\documentclass[aps,prl,showpacs,superscriptaddress,twocolumn]{revtex4}
\usepackage[dvips]{graphicx}
\usepackage{amssymb}

\newcommand{\be}{\begin{equation}}
\newcommand{\ee}{  \end{equation}}
\newcommand{\ba}{\begin{eqnarray}}
\newcommand{\ea}{  \end{eqnarray}}
\newcommand{\bi}{\begin{itemize}}
\newcommand{\ei}{  \end{itemize}}

\begin{document}

\title{Nonlinear conductance in a ballistic Aharonov-Bohm ring}
\author{Alexis R. Hern\'andez}
\affiliation{Laborat\'orio Nacional de Luz S\'{\i}ncrotron, Caixa Postal 6192, 13084-971 Campinas, Brazil}
\affiliation{Centro Brasileiro de Pesquisas F\'{\i}sicas,
             Rua Dr.~Xavier Sigaud 150, 22290-180 Rio de Janeiro, Brazil}
\author{Caio H. Lewenkopf}
\affiliation{Departamento de F\'{\i}sica Te\'orica, Universidade do Estado do Rio de Janeiro,
             20550-900 Rio de Janeiro, Brazil}
             
\date{\today}

\begin{abstract}
The nonlinear electronic transport properties of a ballistic Aharonov-Bohm ring are investigated. It is demonstrated how the electronic interaction breaks the phase rigidity in a two-probe mesoscopic device as the voltage bias is increased. The possibility of studying interference effects in the nonlinear regime is addressed. The occurrence of magnetic field symmetries in higher order conductance coefficients is analyzed. The results are compared with recent experimental data. 
\end{abstract}

%72.10.-d Theory of electronic transport; scattering mechanisms
%73.23.-b Electronic transport in mesoscopic systems

\maketitle
%%%%%%%%%%%%%%%%%%%%%%%%%%%%%%%%%%%%%%%%%%%%%%%%%%%%%%%%%%%%%%%%%%%%%%%%%%%%%%%%%
{\sl Introduction:} 
The Onsager-Casimir \cite{OnsagerCasimir} reciprocity relations are ubiquitous in systems driven out-of-equilibrium in the linear response regime. In classical systems, the quest for related symmetries in the non-linear regime is a subject of increasing interest with potential applications in numerous problems \cite{Andrieux06}.
In mesoscopic physics, the Onsager symmetries enforce, for instance, the linear conductance in two-terminal devices to be an even function with respect to magnetic field inversion \cite{Buttiker86}. Often described as phase rigidity, this prevents the use of such devices for quantum interferometry \cite{Buks96}. Recently, experimental groups have studied how the Onsager-Casimir relations are broken in nonlinear transport as the source-drain bias is increased in a variety of open systems, such as quantum dots \cite{Lofgren04,Zumbuhl06,Marlow06}, carbon nanotubes \cite{Wei05}, and quantum rings \cite{Leturcq06,Angers2007}.
A large bias induces a rearrangement of charges, which are subjected to the Coulomb interaction. Far from equilibrium, interactions break the Onsager-Casimir symmetries \cite{Christen96,Sanchez04,Spivak04}. Theoretical effort has been devoted to infer the effective electron-electron interaction parameters from the statistical analysis of nonlinear transport experiments \cite{Sanchez04}.  

A recent study of nonlinear transport in a two-terminal ballistic Aharonov-Bohm (AB) ring as 
a function of an applied perpendicular magnetic field $B$ and a bias $V$ \cite{Leturcq06} 
unveiled non-statistical puzzling symmetries. By expanding the current $I(B)$ 
in powers of $V$,
\be
\label{eq:generalcurrent}
I \equiv G^{(1)} V + G^{(2)}V^2 + G^{(3)}V^3 + \cdots \;,
\ee
nonlinear conductance coefficients of different orders were analyzed. 
It was observed that the even conductance coefficients $G^{(2n)}$ are neither 
even nor odd with respect to the magnetic field. In contrast, surprisingly, the odd 
ones, $G^{(2n+1)}$, are even functions of $B$ raising the question whether this observation 
is an indication of a new fundamental non-equilibrium symmetry.
Ref.~\cite{Leturcq06} also shows that phase rigidity is lifted in the nonlinear regime. 
The accumulation of an extra phase in one of the quantum ring's arms is mimicked by using 
lateral gates that locally modify the electronic density. 
The phase shift of the AB oscillations in $G^{(2)}$ as a function of lateral 
gates voltages shows a smooth variation with a slope proportional to the 
arm length, which is a promising result towards using AB rings as  
two-terminal interferometers.

In this paper we introduce a simple model that allows for a semi-analytic description of 
the nonlinear transport in a ballistic AB ring. We analyze the nonlinear conductance
oscillation phase shifts and the symmetries of the nonlinear conductance coefficients, both 
issues experimentally addressed in Ref.~\cite{Leturcq06}. We discuss the phase rigidity 
in the $G^{(2)}$ conductance coefficient and present an explanation for the observed even 
parity in the $G^{(3)}$ and $G^{(5)}$ conductance coefficients.

{\sl The method:}
Let us consider the electronic transport in a two-probe mesoscopic ring connected to reservoirs, $\alpha=1,2$, both at a temperature $T$. A voltage bias applied to the reservoirs drives the system out of equilibrium and causes a current flow. In the absence of inelastic processes, the electron current on the lead $\alpha$ reads \cite{Christen96,Buttiker93}
\be \label{eq:compactI}
I_{\alpha}=\frac{2 e}{h}\sum_{\beta=1}^2 \int_{-\infty}^{\infty} dE\;
f_\beta(E) A_{\alpha\beta}(E;\{U({\bf r})\}) \;, \ee
where
$f_\beta(E)=f(E-\mu_0-eV_{\beta})=[e^{(E-\mu_0-eV_{\beta})/k_{\rm B}T}+ 1]^{-1}$,
% is the Fermi function, 
$k_{\rm B}$ is the Boltzmann constant, and $V_\beta$ is the voltage applied to the reservoir 
connected to the lead $\beta$ (measured with respect to the equilibrium chemical potential $\mu_0$). 
Due to current conservation, $I=I_1=-I_2$.

The transmission coefficient $A_{\alpha\beta}$ is given by \cite{Christen96}
\be \label{eq:defA}
A_{\alpha\beta}(E;\{U({\bf r})\})\equiv {\rm Tr} [{\bf 1}_{\alpha}\delta_{\alpha\beta} -
  {\bf S}^\dagger_{\alpha\beta}{\bf S}^{}_{\alpha\beta}]\;,
\ee
where ${\bf S}_{\alpha\beta}$ denotes the scattering matrix with lines (rows) associated with the channels $a$ ($b$) at the 
contact $\alpha$ ($\beta$), ${\bf 1}_\alpha$ is the identity matrix, and the trace runs over all open channels in $\alpha$ 
and $\beta$. The scattering matrix ${\bf S}_{\alpha \beta}(E,\{U({\bf r})\})$ is a function of the electron energy and a functional 
of the electrostatic potential $U({\bf r})$ that is established in the conductor due to the voltages $\{V_\gamma\}$.

It is convenient to expand  $\Delta U({\bf r})\equiv U({\bf r}) - U_{\rm eq}({\bf r})$ as a power series of the applied voltages, namely,
\be
\label{eq:expansionU}
\Delta U({\bf r})= \sum_{\alpha} u_{\alpha}({\bf r})V_{\alpha} + \frac{1}{2}
\sum_{\alpha,\beta}u_{\alpha\beta}({\bf r})V_{\alpha}V_\beta+ \cdots \;,
\ee
with the characteristic potentials $u_{\alpha \beta \cdots}({\bf r})$ defined as
\be
u_{\alpha \beta \cdots}({\bf r})=\left( \frac{\partial}{\partial V_{\alpha}}\frac{\partial}
{\partial V_{\beta}}\cdots\right) U({\bf r}) \Big|_{\{V_\gamma\}=0} . 
\ee
In line with Ref. \onlinecite{Leturcq06}, we take $V_1= V/2$ and $V_2=-V/2$.
% and write the current as in Eq.~(\ref{eq:generalcurrent}).

The linear conductance, $G^{(1)}\equiv \partial I /\partial V|_{V=0}$, is given by the Landauer formula, namely, $G^{(1)} = (2e^2/h) \int dE(-\partial_E f_0) (A_{11}-A_{12})(E;\{U_{\rm eq}({\bf r})\})/2$, where $f_0(E) = f(E-\mu_0)$ and $U_{\rm eq}$ is the equilibrium electrostatic potential for $V=0$. 

Simple symmetry arguments show that $G^{(1)}(B) = G^{(1)}(-B)$. In contrast, there is no general
principle that predicts reciprocity relations for the nonlinear conductance coefficients. 
The inspection of higher powers of $V$ in Eq.~(\ref{eq:generalcurrent})
% and (\ref{eq:expansionU}) 
gives any $G^{(n)}$ in terms of the characteristic potentials and functional derivatives of the transmission coefficients with respect to $U({\bf r})$. While quantities like $\delta A_{\alpha \beta} / \delta U({\bf r})|_{V=0}$ are expressed in terms of ${\bf S}_{\alpha \beta} (E; \{U_{\rm eq}({\bf r})\})$, the $u_{\alpha\beta \cdots}$'s encode the information about electronic many-body interactions \cite{Buttiker93,Hernandez09}.

The conductance coefficient $G^{(2)}$ is written as $G^{(2)}=e^3/2h\int_{-\infty}^\infty dE \left(-\partial_E f_0\right) T^{(2)}$ with  \cite{Christen96}
\ba
\label{eq:I2}
T^{(2)} = \int d{\bf r}\; \big[u_1({\bf r})-u_2({\bf r})\big] \left.
\frac{\delta(A_{11}-A_{12})} {e\delta U({\bf r})}\right|_{V=0}.
\ea
At the simplest approximation level, $u_{\alpha}({\bf r})$ is obtained from a Hartree equation, 
whose source terms are the injected electron density and the corresponding induced charge in 
the conductor, the latter related to the Lindhard function \cite{Buttiker93}. Using the 
Thomas-Fermi approximation for the polarization, one writes \cite{Christen96,Hernandez09} 
\be
\label{eq:Poisson_ulinear}
-\Delta u_{\alpha}({\bf r})+4\pi e^2 u_{\alpha}({\bf r}) \sum_{\beta}\frac{dn^{\rm em}_{\beta}({\bf r})}{dE}=4\pi e^2 \frac{dn^{\rm in}_{\alpha}({\bf r} )}{dE},
\ee
where the injectivity is given by
\be
\label{eq:injec}
\frac{dn^{\rm in}_{\alpha}({\bf r} )}{dE}=\!\int_{-\infty}^{\infty} \frac{dE}{4\pi i} \left(\frac{\partial f_0}{\partial E}\right)\sum_{\beta}{\rm Tr} \left[ {\bf S}^{\dagger}_{\beta \alpha}\frac{\delta {\bf S}_{\beta \alpha}}{e\delta U({\bf r})}-{\rm H.c.}\right]
\ee
and the emissivity, $dn^{\rm em}_{\alpha}({\bf r} )/dE$, is obtained by exchanging $\alpha$ and $\beta$ in ${\bf S}$ at Eq.~(\ref{eq:injec}) \cite{Buttiker93}. Injectivity and emissivities are related by symmetry: ${\bf S}_{\alpha\beta}(B) = [{\bf S}_{\alpha\beta}(-B)]^T$ readily gives that $dn^{\rm in}_{\alpha}(B)/dE=dn^{\rm em}_{\alpha}(-B)/dE$, often called microreversibility relation \cite{Buttiker93}.
Gauge invariance requires that $\sum_\alpha u_\alpha({\bf r}) = 1$, leading to $\sum_{\alpha}dn^{\rm in}_{\alpha}({\bf r})/dE =\sum_{\alpha}dn^{\rm em}_{\alpha}({\bf r} )/dE = dn({\bf r})/dE$, the density of states.

Since $A_{\alpha\beta}$ and $\delta A_{\alpha\beta}/\delta U({\bf r})|_{V=0}$ show the same
symmetries with respect to $B$, one concludes that the characteristic potentials are formally responsible
for the violation of the Onsager relations $G^{(2)}$. The same reasoning holds for higher 
conductance coefficients.

%Using the non-equilibrium Green's function technique, it was shown \cite{Hernandez09} that 
%the $G^{(n)}$ can be formally obtained to arbitrary order.
Expressions for the coefficients $G^{(n)}$ are known to arbitrary order \cite{Hernandez09}.
$G^{(3)}= -e^4/8h \int_{-\infty}^{\infty} dE  (-\partial_E f_0) T^{(3)}$, necessary for what follows, reads
\ba
\label{eq:G3}
T^{(3)} &=&  \int\!  d{\bf r}  \, u^{(2)}({\bf r}) \, \left. \frac{\delta (A_{11}-A_{12})}{e\delta U({\bf r})}\right|_{V=0}+\\
&& 
\!\!\!\!\!\!\!\!\!\!\!\!\int\!  d{\bf r} \int\! d{\bf r'} \left[u^{(1)}({\bf r})u^{(1)}({\bf r'})+\frac{1}{3}\right] \left. \frac{\delta^2 (A_{11}-A_{12})}{e\delta U({\bf r})e\delta U({\bf r'})}\right|_{V=0} \nonumber
\ea
where $u^{(1)}\equiv u_1 - u_2$ and $u^{(2)}\equiv u_{11} + u_{22} - u_{12} - u_{21}$. $u^{(2)}$ is the solution of \cite{Ma98,Hernandez09}
\ba
\label{eq:u12}
\!\!\!\!\!\!\left( \Delta + 4\pi e^2 \frac{dn}{dE} \right) u^{(2)} = &&\!\!\!\!\!\! -4\pi e^3\Big[(u^{(1)})^2 +1\Big]\frac{d^2n}{dE^2}
\nonumber\\
&&\!\!\!\!\!\!-2u^{(1)} \left(\frac{d^2n_1^{\rm in}}{dE^2}-\frac{d^2n_2^{\rm in}}{dE^2}\right) ,
\ea
which does not show any simple universal symmetry. We conclude that the symmetry experimentally observed in $G^{(n>1)}$ are system specific, a subject we investigate next. 

\begin{figure}[ht]
\vskip-0.3cm
\includegraphics[width=4.5cm]{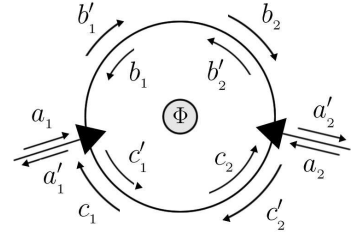}
\vskip-0.3cm
\caption{Two-terminal Aharonov-Bohm ring \cite{Buttiker84}: Arrows show the flow of the electron wave function amplitudes at the vicinity of the contacts, which are indicated by triangles.}
\label{fig:quantum-ring}
\end{figure}

%%%%%%%%%%%%%%%%%%%%%%%%%%%%%%%%%%%%%%%%%%%%%%%%%%%%%%%%%%%%%%%%%%%%%%%%%%%%%%%%%
{\sl Model:} We address the electronic transport through an AB quantum ring using the single-channel model put forward in Ref.~\cite{Buttiker84}. The scattering problem is defined by the electron flow at the vicinity of the contacts as sketched in Fig.~\ref{fig:quantum-ring}. Incoming electron wave function amplitudes are indicated by unprimed Latin indices, while outgoing ones are denoted by primed letters. For instance, $a^{}_\alpha$ ($a^\prime_\alpha$) is the wave function amplitude of an incoming (outgoing) electron at the contact $\alpha$.

%%%%%%%%%%%%%%%%%%%%%%%%%%%%%%%%%%
%%%%%%%%%%%%%%%%%%%%%%%%%%%%%%%%%

The scattering by the contacts connecting the leads to the ring is described by the scattering matrix $S^{(\alpha)}$. Assuming that $(i)$ the effect of the magnetic field on the electrons is negligible on the vertex length scale, $(ii)$ the scattering amplitudes are symmetric with respect to the two branches of the ring, and
$(iii)$ $S^{(\alpha)}$ is real (the complex part of the $S$ matrix can be introduced in the propagation at the arms of the ring); one writes \cite{Buttiker84} 
\be \label{S-vertex} \left(\begin{array}{c}
a'_\alpha \\
b'_\alpha \\
c'_\alpha\\
\end{array}\right)=
\left(\begin{array}{ccc}
-(\chi_\alpha+\xi_\alpha) & \epsilon_\alpha^{1/2} & \epsilon_\alpha^{1/2} \\
\epsilon_\alpha^{1/2} & \chi_\alpha & \xi_\alpha \\
\epsilon_\alpha^{1/2} & \xi_\alpha & \chi_\alpha \\
\end{array}
\right)\left(
\begin{array}{c}
a_\alpha \\
b_\alpha \\
c_\alpha\\
\end{array}
\right)
\ee
with $\chi_\alpha=-(\sqrt{1-2\epsilon_\alpha}-1)/2$ and $\xi_\alpha=-(\sqrt{1-2\epsilon_\alpha}+1)/2$.
The parameter $\epsilon_\alpha \in [0, 1/2]$ tunes the reflection at the vertex $\alpha$. The contact is closed for $\epsilon_\alpha =0$ and maximally open for $\epsilon_\alpha =1/2$.
When $S^{(1)}=S^{(2)}$ the contacts are identical and the ring has, by construction, a reflection symmetry.

The electron propagation is described by a transfer matrix which, for the upper arm, renders $b_2 = e^{i(k\ell_1-\phi_1)}b_1'$ and $b_2'= e^{-i(k\ell_1+\phi_1)} b_1$,
where $k$ is the electron wave vector, $\ell_{1}$ is the arm length, and $\phi_{1}$ is the AB phase accumulated by the electron while traversing the ring from vertex 1 to 2. A similar transfer matrix can be written for the amplitudes $c$ and $c'$ by replacing $\ell_1 \rightarrow \ell_2$ and $\phi_1 \rightarrow -\phi_2$. The full scattering matrix $S_{\alpha\beta}$ is readily obtained from  $S^{(\alpha)}$ and the above described transfer matrices.

The electronic interactions are accounted for in  $\delta S_{\alpha\beta}/ \delta U({\bf r})$ as follows. We place a $\delta$-function scatterer at the position $\bf r$ along the ring, calculate the modified transfer matrices, and take the limit
\be
\frac{\delta S_{\alpha\beta}}{\delta U({\bf r})}=\lim_{\Omega\rightarrow 0} \frac{S_{\alpha\beta}[U({\bf r}')+ \Omega \delta ({\bf r}'-{\bf r})]-S_{\alpha\beta}[U({\bf r}')]}{\Omega}\;.
\ee
$\delta S_{\alpha\beta}/\delta U({\bf r})$ is used to compute injectivities, emissivities, and $\delta A_{\alpha\beta}/\delta U({\bf r})$. 

To relate our numerical findings to experiments, we choose the quantum ring diameter and set the effective electron mass to keep about 50 electrons below the Fermi energy. We present results for zero temperature. 

\begin{figure}[h!]
\vskip1.7cm
\includegraphics[width=9.0cm]{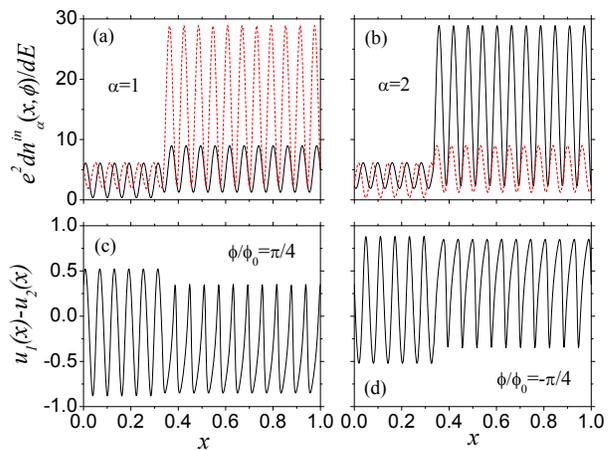}
\vskip-2.8cm
\caption{ Injectivity at $\alpha=1$ (a) and $\alpha = 2$ (b) for $\phi/\phi_0=\pm \pi/4$. The
variable $x\in [0,1]$ describes the position along the ring starting at the contact 1. Contact 2 is placed at $x=1/3$. Characteristic potential,  $u^{(1)}(x)$ for a symmetric bias and magnetic flux (c) $\phi/\phi_0=
\pi/4$ and (d) $\phi/\phi_0= -\pi/4$.} \label{injectivity}
\end{figure}

{\sl Results:} Figures \ref{injectivity}(a) and \ref{injectivity}(b) illustrate typical injectivities calculated with our model. They present discontinuities at the contact positions, oscillations of periodicity about $1/k_F$ around plateaus of different amplitudes for each arm. Oscillations in $dn^{\rm in}_{\alpha}(B)/dE$ are Friedel-like fringes due to trapped charges coming from contact $\alpha$ while the plateaus are related to the charge flow also coming from contact $\alpha$. By computing the emissivity we verify microreversibility $dn^{\rm in}_{\alpha}/dE(B)=dn^{\rm em}_{\alpha}/dE(-B)$. We also check that $\sum_\alpha dn^{\rm in}_\alpha/dE(B)=\sum_\alpha dn^{\rm em}_\alpha/dE(B)$.

%%%%%%%%%%%%%%%%%%%%%%%%%%%%%%%%%%
%%%%%%%%%%%%%%%%%%%%%%%%%%%%%%%%%

Since our primary goal is a qualitative understanding, we do not attempt to microscopically model the screening effects of a specific device. Instead, we use the contact interaction approximation and algebraically solve Eq.~(\ref{eq:Poisson_ulinear}) to write the characteristic potentials as
\be \label{eq:potcont} u_{\alpha}=\frac{4\pi e^2
dn_{\alpha}^{\rm in}/dE}{\kappa^2 + 4\pi e^2 dn/dE} , \ee
where $\kappa$ is a constant that characterizes the electron interaction strength. In the one-dimensional case considered here, $\kappa$ is dimensionless. We set $\kappa=0.1/(k_F \ell)$, with $\ell = \ell_1 + \ell_2$. Illustrative characteristic potentials are shown in Figs.~\ref{injectivity}(c) and (d).

%%%%%%%%%%%%%%%%%%%%%%%%%%%%%%%%%%
\begin{figure}[h!]
\includegraphics[width=8.cm]{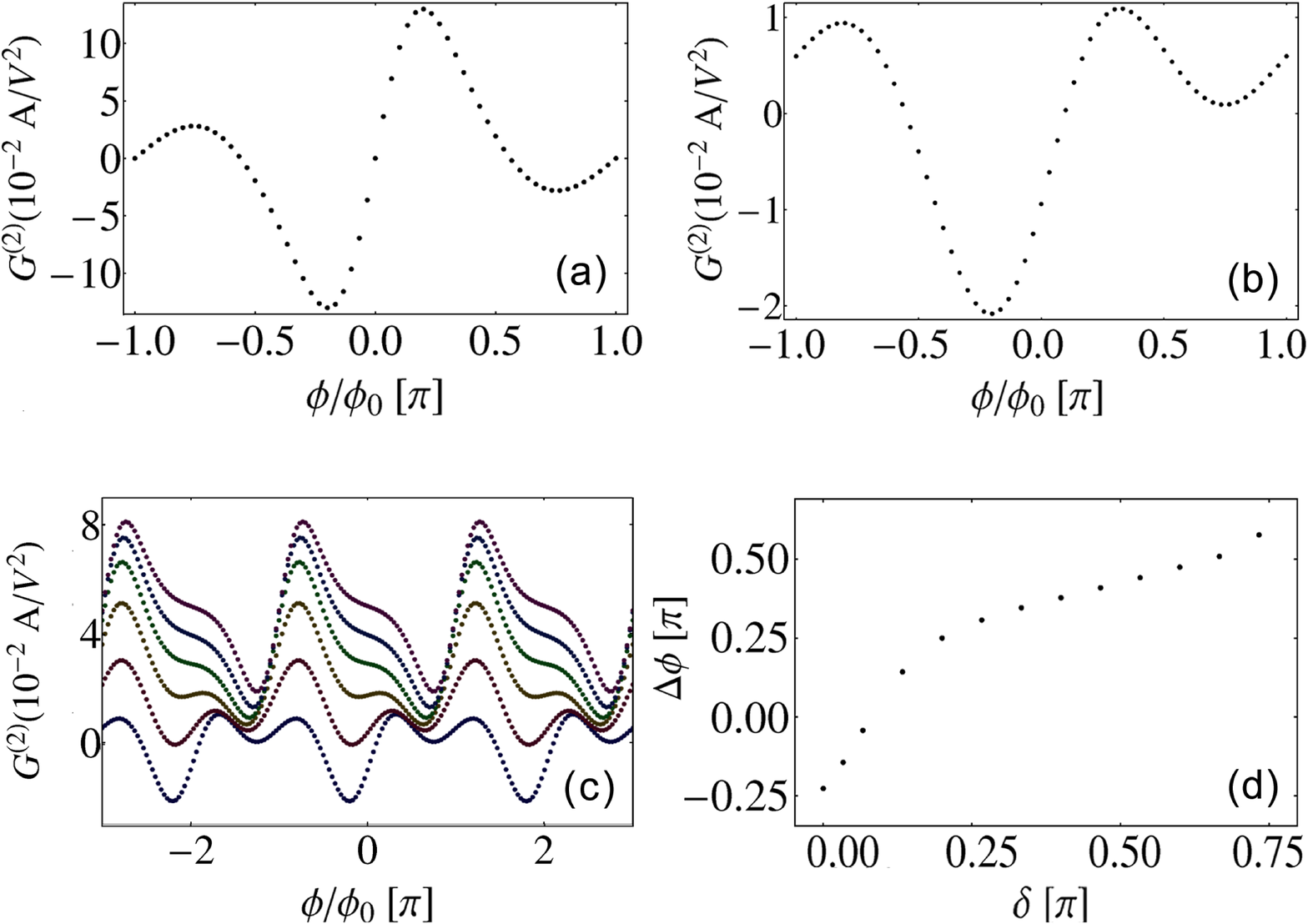}
\caption{\label{NLCGraf} Nonlinear conductance $G^{(2)}$ as a function of $\phi/\phi_0$ for both 
(a) a symmetric ring and (b) an asymmetric ring. (c) Nonlinear conductance $G^{(2)}$ for an
asymmetric ring as a function of magnetic flux for different values of $\delta=\Delta k \ell_{i}$. An offset was introduced to separate the curves. (d) Phase of the first harmonic of $G^{(2)}$ as a function of the phase $\delta$.}
\end{figure}
%%%%%%%%%%%%%%%%%%%%%%%%%%%%%%%%%

Inserting $u_\alpha({\bf r})$ and $\delta A_{\alpha \beta}/\delta U({\bf r})$ in Eq.~(\ref{eq:I2}) 
we obtain $G^{(2)}$. Figure \ref{NLCGraf}(a) shows $G^{(2)}$ versus magnetic flux $\phi$. Due to 
the system spatial symmetry introduced by taking $S^{(1)}=S^{(2)}$, $G^{(2)}$ is odd in 
magnetic field. This peculiar nonlinear phase rigidity disappears as the spatial symmetry is 
broken. To realize this, we transform the scattering matrix of $\alpha=2$ as
$S^{(2)}\rightarrow R_x^{-1}(\theta)S^{(2)} R_x^{}(\theta)$, where $R_x(\theta)$ is the matrix
representation of a classical rotation by $\theta$ around the $x$-axis.
We observe that as $\theta$ increases the symmetry of $G^{(2)}$ is gradually broken.
Fig.~\ref{NLCGraf}(b) shows $G^{(2)}$ for $\theta =\pi/6$.  

Phase locking in two-terminal devices is lifted as the source-drain bias is increased. The phase
shift in the AB oscillations of $G^{(2)}$ were experimentally studied by controlling 
the electron density at a selected arm \cite{Leturcq06}. 
We model this by changing the wave vector at one of quantum ring arms, namely, $k\rightarrow k+
\Delta k$. This is equivalent to increasing the accumulated phase of the electrons flowing through
this arm by $\delta =\Delta k \,\ell_{i}$. Figure \ref{NLCGraf}(c) shows the conductance $G^{(2)}$
for several values of $\delta$.

\begin{figure}[ht]
\includegraphics[width=6cm]{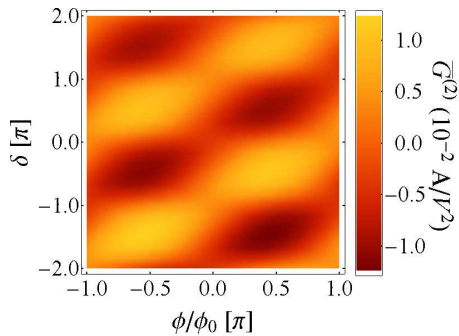}
\caption{\label{Fig:NLC3D} [Color online] $\bar{G}^{(2)}$, the first harmonic of $G^{(2)}$, as a function of magnetic
flux $\phi/\phi_0$ and $\delta$.}
\end{figure}

In experiments, dephasing suppresses the higher harmonic contributions to $G^{(2)}$ 
that correspond to electron trajectories that go around the ring more than once. 
Since our treatment does not include decoherence effects we extract the first harmonic,
$\overline{G}^{(2)}$, of $G^{(2)}$ to interpret the experimental data. Using the formula
$\tan(\Delta\phi)=\langle G^{(2)} (\phi)\sin(\phi/\phi_0)\rangle / 
\langle G^{(2)}(\phi)\cos(\phi/\phi_0) \rangle$ \cite{Leturcq06} we evaluate $\Delta\phi$ 
for different 
values of $\delta$. Figure \ref{NLCGraf}(d) collects the values corresponding to the curves 
shown in Fig.~\ref{NLCGraf}(c). 
The dependence of $\overline{G}^{(2)}$ on $\delta$ and the magnetic flux $\phi/\phi_0$ is shown 
in Fig.~\ref{Fig:NLC3D}.

%%%%%%%%%%%%%%%%%%%%%%%%
%%%%%%%%%%%%%%%%%%%%%%%%

In Fig.~\ref{NLCGraf}(d) we observe a smooth monotonic dependence between $\Delta\phi$ and 
$\delta$. Although the dependence is not linear, except for small $\delta$, it is in accordance 
with the experiment \cite{Leturcq06}. 
However, since the perturbation $\delta$ modifies both, the characteristic potential and the
functional derivative of $A_{\alpha \beta }$ in a non-trivial way, a monotonic behavior is 
not generic.  Varying the model parameters, we also find cases where $\Delta\phi$ shows fast jumps as well as non-monotonic behavior as a function of $\delta$. We conclude that despite eliminating the phase rigidity by applying a large bias, it is still a difficult task to use the nonlinear transport regime for interferometry studies. 

We now turn our attention to the parity of $G^{(n)}$ with respect to $B$. We already concluded that, in general, $G^{(3)}$ is neither even nor odd in $B$. Let us present a system specific scenario that explain the experimental observations \cite{Leturcq06}:
Most inelastic processes are very weakly dependent on the magnetic field. For a large bias, those
give even parity contributions to the $I-V$ characteristics. Hence, we focus on the odd parity contributions.  
When the contacts are equivalent, the spatial symmetry imposes that $u_1(B,{\bf r})=u_2(-B,{\bf r})$ and the quantum coherent part of $G^{(2)}$ is odd in $B$. Similar arguments, applied to $u^{(2)}$ in Eq.~(\ref{eq:u12}), lead to a purely even $G^{(3)}$. In this scenario, inelastic processes even in $B$ explain why $G^{(2)}$ does not have a definite parity and $G^{(3)}$ has. The fact that the ring is quite open, favors the hypothesis of nearly equivalent contacts. 
Further support for this picture comes from the analysis of the $h/2e$ oscillations observed at low magnetic fields \cite{Leturcq06}. Those can be attributed to paths going twice around the ring or to time-reversed paths each traversing the ring only once \cite{AAS}. While the first ones have no define $B$ parity, the latter cause the Altshuler-Aronov-Spivak oscillations, which are usually more robust and by nature even functions in $B$. Hence, the absence of AAS 
oscillations in even part of the measured $G^{(2)}$ and $G^{(4)}$ is consistent with the hypotheses of a nearly symmetric ring.

In summary, we presented a theoretical approach that allows for a semi-quantitative discussion of nonlinear transport properties of ballistic mesoscopic devices, in particular of AB rings. We described how phase rigidity is broken in two-terminal devices as the voltage bias is increased. We also found that the experimentally observed symmetry $G^{(2n+1)}(B)=G^{(2n+1)}(-B)$ is not generic and can be explained in terms of systems specific features. Further experiments, exploring other control handles and geometries, can greatly contribute to settle this problem.  

We thank M. B\"uttiker, K. Ensslin, and R. Leturcq for valuable discussions and acknowledge financial support from the Brazilian funding agencies FAPERJ and CNPq.

{\it Note added:} Related phenomena were very recently reported on a different system: An AB-ring with discrete level spectrum due to an embedded quantum dot (QD)\cite{Meir09}. Here, the weak tunneling strengths make the asymmetry in the $u$'s much weaker than in our paper and many-body effects in the QD can play a dominant role.

%%%%%%%%%%%%%%%%%%%%%%%%%%%%%%%%%%%%%%%%%%%%%%%%%%%%%%%%%%%%%%%%%%%%%%%%%%%%%%%%%%%%%%%%

\end{document}